\documentstyle[12pt]{article}
\normalbaselines
\def\ii{\'\i}

\begin{document}
{}~~
\vskip .35in
\begin{center}
{\LARGE\bf Pseudo-Spin Symmetry and its Applications} \footnote{ \small Work
supported in part by
project UNAM-CONACYT 3513-E9310. }\\[1pc]
{\large Octavio Casta\~nos$^{*}$, Jorge G. Hirsch$^{\dagger}$ and Peter O.
Hess$^{*}$    }\\[1pc]
{\small \it $^{*}$ Instituto de Ciencias Nucleares, UNAM; A. Postal 70-543,
04510}  \\
{\small \it  $^{\dagger}$ Departamento de F\ii sica, CINVESTAV; A. Postal
14-740,  07000}\\
{\small \it   M\'exico, D. F.}
\end{center}

\vspace{2.0truepc}
\vbox{\rightskip=2truepc
\hangindent = 2truepc
\hskip 2truepc  \small {\bf Abstract}. The pseudo-spin symmetry is reviewed. A
mapping that produces the separation of the total angular momentum into
pseudo-orbital and pseudo-spin degrees of freedom is discussed, together with
the analytic transformations that take us from the normal parity space to the
eigenstates of a pseudo-oscillator with one quanta less. The many-particle
version of the unitary transformation to the pseudo-SU(3) space is established.
As an example, these symmetries are used to describe the double beta decay
phenomenon in heavy deformed nuclei.}

\vspace{2.0truepc}

\centerline{\large \bf INTRODUCTION}

\vspace{1truepc}
The pseudo-spin symmetry was introduced in [1,2], where a separation of the
single particle Hilbert space into pseudo orbital and pseudo-spin degrees of
freedom was proposed. This new coupling scheme provides a simple interpretation
of many striking features of the Nilsson orbits: i) The observed approximate
degeneracy of the single particle levels of the type $(N, l, s) j$ and $(N, l +
2, s) j+1$ for the spherical case or the one associated to the levels with
asymptotic quantum numbers $[N, N_z, \Lambda ] \,  \Omega = \Lambda + {1 / 2}$
and $[N, N_z, \Lambda + 2 ] \, \Omega = \Lambda + {3 / 2}$ for the large
deformation case. These energy orbitals can be considered as pseudo-spin orbit
doublets, implying that the strength of the pseudo spin-orbit interaction is
small. ii) The expectation value of the pseudo-spin operator, iii) the
calculation of the decoupling parameters, and iv) the matrix elements of the
Coriolis interaction [1]. More recently the pseudo-spin symmetry has been
considered to be useful to characterize identical bands [3] and as a signature
of superdeformation [4].

For heavy nuclei the nuclear shells can be divided into two parts: the
associated to the single particle orbits of the same parity, which are called
normal parity levels and the intruder that comes from the shell above of
opposite parity and thus it has been called unique or abnormal parity orbital.
Thus for a given shell, $N_\alpha$ one has
$ j_\alpha^N  =  \{ {1 / 2}, \,  {3 / 2}, \, \cdots \, ,  N_\alpha - { 1 / 2}
\, \}$,
$ j_\alpha^A  =  N_\alpha + {3 /2}$, where $\alpha = \pi \, \hbox{or} \,  \nu$
for protons and neutrons, respectively.
This separation of the space for the single particle states of Nilsson or Wood
Saxon potentials is reasonable because it has been proved that the intruder
orbitals are nearly unmixed for standard deformations [5].
When relabelled in the pseudo spin scheme, the normal parity levels form a
major shell for a pseudo-oscillator potential with $ \tilde N_\alpha = N_\alpha
- 1$ quanta. Since the symmetry of this oscillator is of course SU(3), one may
consider the pseudo-SU(3) coupling scheme.

In the Second Section, we review the transformation to the pseudo space
introduced by Bohr, Mottelson and Hamamoto (BHM) [6] and the mappings from
oscillators to pseudo-oscillators wave functions for the spherical and
asymptotic cases [7].  We emphasize the difference between the pseudo-spin and
the pseudo-oscillator symmetries. The first one implies the separation of the
normal parity degrees of freedom into pseudo spin and pseudo orbital spaces
(called pseudo-space) while the second one for the one particle case is related
with the use of oscillator wavefunctions with spin to describe the
pseudo-space. At the end of this section, the many-particle unitary
transformation to the pseudo-SU(3) space is introduced and applied to one body
operators. In the final section, theoretical calculations for the $2\nu$ double
beta decays of $^{150}$ Nd and $^{238}$U are presented [8].

\vspace{2truepc}

\centerline{\large \bf UNITARY TRANSFORMATIONS}

\vspace{1truepc}

The BHM unitary transformation acts on the angular-spin parts of the wave
functions of the particle, without affecting the radial motion. It is given by
the scalar product between the spin operator $\vec{S}$ and $\hat{{r}}$ a unit
vector in the direction of the position of the particle, $U_{BHM} = 2 \, i \,
\hat{{r}} \cdot \vec{S}$. It is  straightforward, to calculate the action on
the single-particle wave function of the three dimensional harmonic oscillator
with quantum numbers
\begin{equation}
\psi_{ N ( l,s) jm}(\vec{r}) = R_{N l} (r) \sum_{\mu \sigma}{\langle l \mu, s
\sigma | j m\rangle} Y_{l \mu}(\theta, \phi) \chi_{s \, \sigma} \, ,
\end{equation}
which can be written as a product of the radial part times the superposition of
two spinor wave functions [6]. This transformation gives rise to the results $l
= j \pm {1/2} \ \mapsto \ \tilde l = j \mp {1/ 2} \, $. Thus it produces the
breaking of the normal degrees of freedom into pseudo-spin and pseudo-orbital
spaces, but it does not give the wave functions of a pseudo-oscillator with one
quanta less.

In the set of wave functions (2) corresponding to a given $N$ value, we
disregard those with $j = N + {1 \over 2}$, which define the unique or abnormal
space, and only consider the remaining wave functions of the normal parity
orbitals. These states can be mapped onto the eigenstates $ | \tilde N (\tilde
l \tilde s) \tilde j \tilde m \rangle$ of a pseudo-oscillator via the following
unitary operator [7]
\begin{equation}\label{a}
U = 2 \, ( \vec{\xi} \cdot \vec{S}) \,( \hat N - 2 \vec{L} \cdot \vec{S} )^{-{1
\over 2}} \, ,
\end{equation}
where $\vec{\xi}$ is the annihilation harmonic oscillator quantum. The action
of $U$ onto eigenstates of a three dimensional harmonic oscillator gives the
result
\begin{equation}
U | N ( l,s) jm \rangle = | \tilde N (\tilde l \tilde s) \tilde j \tilde m
\rangle \, ,
\end{equation}
with the relations between the labels $\tilde N = N - 1, \, \tilde s = s, \,
\tilde j = j, \, \tilde m = m$, and $\tilde l = l \pm 1 \hbox{ according to
whether} \, j = l \pm { 1/2} \ $. Although the previous mapping connecting the
normal parity harmonic oscillator eigenstates with the full set of eigenstates
of a pseudo-oscillator was proposed long ago [1, 2], the explicit form of the
unitary operator (3) was constructed for the first time in [7]. More recently,
the transformation to the pseudo-oscillator space has been carried out by using
the Algebraic Generator Coordinate Method [9].

Now we consider the asymptotic wave functions of the Nilsson Hamiltonian [7]
associated with large deformations. Then the Nilsson orbitals are described by
the states characterized by $| \, [N, \, N_z, \, \Lambda] \, \Omega \,
\rangle$, the cylindrical harmonic oscillator states with spin. The degeneracy
mentioned above can be explained through the unitary transformation
\begin{equation}
U_{\infty} = 2 \, ( \xi_{+} S_{+} + \xi_{-} S_{-} ) \, ( \hat N_{\rho} - 2
\vec{L}_{z} \cdot \vec{S}_{z} )^{-{1 \over 2}} \, ,
\end{equation}
where the operators $\xi_{\pm}$ and $\eta_{\pm}$ are the spherical components
of the creation $\vec{\eta}$ and annihilation $\vec{\xi}$ harmonic oscillator
operators, the spin operators are expressed in terms of the Pauli matrices $
S_{m} = \sigma_{m}/2$ and $\hat N_\rho$ is the number operator  in the plane.

The $U_{\infty}$ and their corresponding hermitean conjugated operator $
U^{\dagger}_{\infty} $ are unitary, if they act onto states belonging to the
normal space orbitals, that is if we disregard the levels of a given $N$ shell
that satisfy $\Omega = N_\rho + { 1/ 2}$. Thus, the unitary transformations for
the spherical and asymptotic Nilsson orbitals to the pseudo-oscillator wave
functions are not equivalent. This fact which has not been previously
emphasized, is reflected in the separation itself into normal and abnormal
parity spaces. However for standard deformations the exact wave functions of
the Nilsson Hamiltonian can be evaluated and it is observed that there is no
mixing of the unique (spherical) parity orbital with the remaining levels of
the shell [10], which suggest that for these deformations the spherical
transformation is the most appropriate. The division of the shell model space
into normal and unique as proposed in the asymptotic limit may be useful in
applications to superdeformation phenomena.

For a system of $n$ particles the unitary transformation to the
pseudo-oscillator space is defined by
\begin{equation}\label{a1}
U = \prod_{s}^{n}{ U_s} \, .
\end{equation}

In the Fock space, if we denote by $\hat U$ the corresponding unitary operator
associated to $U$, then a general one body operator $F$ is mapped to the
pseudo-SU(3) space by $\hat{ \bar F} = \hat U \hat F \hat U^{\dagger}$. If two
single particle state vectors $|\alpha \rangle$ and $|\tilde\alpha \rangle$
are related by a unitary transformation $U$
(see Eq.~\ref{a}), then the fermion operators are related by $\hat U
a^{\dagger}_\alpha \hat U^{\dagger} = a^{\dagger}_{\tilde{\alpha}}$. Then, $F$
in the pseudo-space is given by
\begin{equation}
\hat{ \bar F} = \sum_{N, \, l, \, j, \, \mu} {^{'} \ \sum_{N', \, l', \, j', \,
\mu'}{ ^{'} \  \langle N' \, (l', {1 \over 2}) \, j' \, \mu' | F_1 |  N \, (l,
{1 \over 2}) \, j \, \mu \rangle} \,
a^{\dagger}_{\tilde N', \,\tilde l', \, \tilde j', \, \tilde \mu'} \ a_{\tilde
N, \,\tilde l, \, \tilde j, \, \tilde \mu}} \, ,
\end{equation}
where the primes on the sum indicate that the unique parity orbitals are
excluded. A similar result is obtained for the two body operators, that is only
the labels of the fermion creation and annihilation operators are changed by
its pseudo quantum numbers. The previous results give formal support to the
recipe indicated in Refs. [1, 11]. The many body expressions in the pseudo
space of the $SU(3)$ generators are given by the series expansions
\begin{equation}
\hat{ \bar L}_q  =  k_L \ \hat{ \tilde L}_q \ + \cdots \, , \qquad
\hat{ \bar Q}_\mu  =  k_Q \ \hat{ \tilde Q}_\mu \, + \cdots  \, ,
\end{equation}
where only the leading terms of the series are indicated. These operators $
\hat{ \tilde L}_q$,$\hat{ \tilde Q}_\mu \, $ have the same form as the $SU(3)$
generators but of a shell with one quanta less, but we want to emphasize that
they are not the transformed operators $\hat{ \bar L}_q$,$\hat{ \bar Q}_\mu$ to
the pseudo space.

\vspace{2truepc}

\centerline{\large \bf DOUBLE BETA DECAY}

\vspace{1truepc}

Double beta decay is a rare transition between two nuclei with the same mass
number $A$ involving change of the nuclear charge number $Z$ by two units. This
exotic phenomenon is a useful tool to test the lepton number conservation,
neutrino properties and models of nuclear structure [12] . It can be classified
into various modes according to the
light particles besides the electrons associated with the decay. The two
neutrino mode ($\beta \beta_{2\nu}$), in which two electrons and two neutrinos
are emitted, takes place independently of the neutrino properties, and
conserves the electric charge and lepton number. The $0\nu$ mode
violates lepton number conservation and therefore it is forbidden in the
standard electroweak theory. To proceed the $0\nu$ decay, the virtual neutrino
must be emitted in one vertex and absorbed in the other one, thus it is
required that: i) the exchanged neutrino is a Majorana particle ($\nu = \bar
\nu)$ and ii) both neutrinos have a common helicity component [13].

Next, we restrict to describe, within the pseudo-SU(3) formalism, the $2\nu$
double beta decays of $^{150}$Nd and $^{238}$U. The decay rate of the
$2\nu$-mode can be calculated through the formulae
\begin{equation}
(\tau^{1/2}_{2\nu})^{-1} = G_{2\nu} \ | \ M_{2\nu}^{GT} \ |^2 \,  ,
\end{equation}
where $G_{2\nu}$ is a kinematic factor and $M_{2\nu}^{GT} = M_{2\nu}$ is a
nuclear matrix element strongly dependent on the considered nuclear model, that
is
\begin{equation}
M_{2\nu} = \sum_{N}{}{1\over{E_0 + E_N -E_i}} \langle  0^+_f \ || \, \Gamma \,
|| \ 1^+_N \rangle \  \langle1^+_N \
 || \,\Gamma \, ||\  0^+_i \rangle \, .
\end{equation}
The $\Gamma$ is denoting the Gamow-Teller operator
$ \Gamma_m = \sum_s \ \sigma_{ms} t^-_s $ and the $E_0= {1 \over 2} {\it
Q}_{\beta  \beta}+ m_e c^2$  is the half of the total energy
released. The $E_N$ gives the energy of the intermediate state $| 1^+_N
\rangle$. The $E_i$ is the energy of the ground state of the initial nucleus $
\vert 0^+_i \rangle$ and the ket
$| 0^+_f \rangle$ describes the ground state of the final nucleus.

In order to compute $M_{2\nu}$, we have to perform a sum over all the
intermediate states . Fortunately, an alternative
form of calculate this matrix element has been developed  [14], {\it i.e.},
\begin{equation}
 M_{2\nu} ={1 \over E_0 }  \langle 0^+_f| \sum_{m}{}
                (-1)^m \Gamma_{-m} {\it F}_m | 0^+_i \rangle,
\end{equation}
where the operator $ {\it F}_m $ is defined by:
\begin{equation}
{\it F}_m = \sum_{\lambda}^{\infty}{{(-1)^{\lambda}
                \over E^{\lambda}_0} \ [ H, [ H, \ldots,[ H,
                \Gamma_m] \ldots ]^{(\lambda -times)} \  .}
\end{equation}
A reasonable model for
describing spectra and $BE2$ transitions of heavy deformed nuclei is [8]:
\begin{equation}
H = \sum_{\alpha}{H_\alpha - {1 \over 2} \chi \  \bf{Q}^a \cdot \bf{Q}^a} +
\zeta_1 \; K^2 + \zeta_2 \; L^2,
\end{equation}
where $H_\alpha$ denotes the spherical Nilsson Hamiltonian for neutrons or
protons plus a constant term $V_\alpha$, which represent the depth of the
potential well. The quadrupole-quadrupole interaction in a given shell and
$K^2$  is a linear combination of  ${L}^{2}$,
${X}_{3}$ and ${X}_{4}$, which are rotational scalar operators built with
generators of the algebra of $SU(3)$ [15].

Notice that the
quadrupole-quadrupole force, $L^2$ and the $ K^2$ interaction are independent
of the
spin degrees of freedom and symmetric in the neutron and proton components. To
evaluate (13), we express $H_{\alpha} $ and $\Gamma_m$ in the second
quantization formalism,
\begin{eqnarray}
H_{\alpha} & = & \hbar \omega \sum_{\eta \, l \, j \,m}
{\epsilon_{\alpha}    \,(\eta,l,j) \  a^{\dagger}_{\eta l
                \, {1 \over 2} j m \alpha}a_{\eta l \, {1\over 2} j m \alpha}}
\, , \\                       \Gamma_m & = &  \sum_{\pi \, \nu}{ \sigma (\pi
,\nu ) A(\pi ,\nu, m)} \, ,
\end{eqnarray}
with $A(\pi ,\nu, m) = [a^{\dagger}_{\eta_{\pi} l_{\pi} \, {1 \over 2}
;j_{\pi}} \bigotimes{\tilde a_{\eta_{\nu} l_{\nu} \, {1 \over 2};
j_{\nu}}}]^1_m$ denoting the angular coupling of proton creation and neutron
annihilation operators, $\epsilon_{\alpha}$
the single particle energies and
\begin{equation}
\sigma (\pi ,\nu ) \equiv  \sum_{\pi \, \nu} {
\sqrt{{2 j_\pi +1 \over 3}} \  \langle \eta_{\pi} l_{\pi} \,
{1 \over 2} \, ; j_{\pi} || \sigma || \eta_{\nu} l_{\nu} \, {1 \over 2} \,;
j_{\nu} \rangle } \, .
\end{equation}
Afterwards some algebraic manipulations, $M_{2\nu}$ is rewritten by
\begin{eqnarray}
M_{2\nu}  = & \sum_{\pi \, \nu} \sum_{\pi {\, ' }\, \nu{\, ' }}
                \sigma (\pi{\, ' } ,\nu{\, ' } ) \ \sigma (\pi ,\nu ) \
                { \langle 0^+_f \ | \ \vec A \, (\pi{\, ' }, \, \nu{\, ' } )
\cdot \vec
		A \, (\pi, \, \nu )  \ | \ 0^+_i \rangle
                \over E_{0} \ + \ \epsilon\,(\eta_{\pi}, \, l_{\pi}, \,
j_{\pi}) \
                 -  \ \epsilon\,(\eta_{\nu}, \, l_{\nu}, \, j_{\nu})} \,
\end{eqnarray}
In the pseudo SU(3) Hilberty space the above sum is restricted to
$\eta_{\pi}=\eta_{\nu}$,
$l_{\pi}=l_{\nu}$, $\eta_{\pi {\, ' }}=\eta_{\nu {\, ' }}, \, l_{\pi {\, '
}}=l_{\nu {\, ' }}, \, j_\pi = j_\nu + 1 $. Following [8], the single particle
energy difference in the denominator takes the form
\begin{equation}
 \epsilon (\eta ,l,j_\pi ) - \epsilon (\eta ,l,j_\nu) = -\hbar \omega k_\pi 2
j_\pi + \Delta_C \, .
\end{equation}
The constants $k_\alpha $ are well known [16] and $\Delta_C$ is used to
determine the value $V_\nu - V_\pi$. The $\Delta_C$ is the difference Coulomb
energy and it is evaluated by the expression
\begin{equation}
\Delta_C ={ 0.70 \over A^{1/3}} [2 Z + 1 - 0.76 ( (Z+1)^{4/3} -Z^{4/3} )]
  MeV \, .
\end{equation}

The description of the correlated deformed ground states is done using
the pseudo SU(3) scheme for the normal parity space and seniority zero
configurations, that is all the nucleons coupled by pairs to angular momentum
zero, for the unique part [11, 15]. The occupancies of these spaces are
determined from the corresponding Nilsson diagrams by selecting a reasonable
deformation and filling each level with a pair of particles in order of
increasing energy. These numbers fix the totally antisymmetric irreducible
representations (irreps) of the unitary groups associated to the normal $U(
(N_\alpha + 1) (N_\alpha + 2)) $ and unique $U( 2 N_\alpha + 4)$ spaces.

For the normal parity space, one separates the degrees of freedom in
pseudo-orbital $U(\Omega_\alpha^N )$ and pseudo-spin $U_\alpha(2)$ parts, with
$ \Omega_\alpha^N = (N_\alpha + 1) (N_\alpha + 2)/2$ and their irreps $\{\tilde
f_\alpha\}$ are characterized by the partitions of the number of particles in
the normal part, $n^N_\alpha$. Then of the pseudo $SU(3)$ irreps,
$(\lambda_\alpha,
\mu_\alpha)$, contained in $\{\tilde f_\alpha\}$, one considers those with
maximum
eigenvalue of the Casimir operator, $(C_2)_\alpha = (\lambda_\alpha +
\mu_\alpha +3) ~ (\lambda_\alpha + \mu_\alpha) - \lambda_\alpha \mu_\alpha$.
In Table 1, the results found for the participant nuclei in the double beta
decays of $^{150}$Nd and $^{238}$U are presented.

\vspace{1pc}
{\bf Table 1.}  Ground states in the Pseudo-SU(3) coupling scheme.

\begin{tabular}{c c c c c c c c c}
\hline
\noalign{\vskip .1truecm }
NUCLEUS& $n_\pi^N$  & $n_\pi^A$ & $n_\nu^N$ & $n_\nu^A$ &  $U(\Omega_\pi^N )$ &
$U(\Omega_\nu^N )$ & $SU_\pi (3)$ & $SU_\nu (3)$ \\
\noalign{\vskip .1truecm }
\hline
\noalign{\vskip .1truecm }
$^{150}$ Nd & $6$ & $4$ & $6$ & $2$ & $\{ 2^3 \}$ & $\{ 2^3 \}$ &
$(12, \, 0)$ & $(18, \, 0)$   \\
\noalign{\vskip .1truecm }
\noalign{\vskip .1truecm }
$^{150}$ Sm & $6$ & $6$ & $4$ & $2$ & $\{ 2^3 \}$ & $\{ 2^2 \}$ &
$(12, \, 0)$ & $(12, \, 2)$   \\
\noalign{\vskip .1truecm }
\noalign{\vskip .1truecm }
$^{238}$ U & $6$ & $4$ & $12$ & $8$ & $\{ 2^3 \}$ & $\{ 2^6 \}$ &
$(18, \, 0)$ & $(36, \, 0)$   \\
\noalign{\vskip .1truecm }
\noalign{\vskip .1truecm }
$^{238}$ Pu & $6$ & $6$ & $10$ & $8$ & $\{ 2^3 \}$ & $\{ 2^5 \}$ &
$(18, \, 0)$ & $(30, \, 4)$   \\
\noalign{\vskip .1truecm }
\hline
\end{tabular}

{\bf Table 2.}  Theoretical estimates for the half-life
$\beta\beta$-decay in the $2\nu$ mode.

\begin{tabular}{c c c c c c c c c}
\hline
\noalign{\vskip .1truecm }
$Transition$  &  $< 0^+_f | \Gamma^2 | 0^+_i >$  & $E[MeV]$ &
$\tau^{1/2}_{theo} [y]$ & $\tau^{1/2}_{exp} [y] [13]$ \\
\noalign{\vskip .1truecm }
\hline
\noalign{\vskip .1truecm }
$^{150}Nd \to ^{150}Sm $  &  $1.31$& $12.2 $ & $6.0 \times 10^{18}$   & $9$-$17
\times 10^{18}$ \\
\noalign{\vskip .1truecm }
\noalign{\vskip .1truecm }
$^{238}U \to ^{238}Pu$  &  $1.51$ & $16.8$ & $1.4 \times 10^{21}$ & $ 2 \times
10^{21}$   \\
\noalign{\vskip .1truecm }
\hline
\end{tabular}

\vspace{1pc}

Finally we used the strong coupled limit [11],  and from the Kronecker
product $
(\lambda_\pi, \mu_\pi) \times (\lambda_\nu, \mu_\nu)$,
the $(\lambda_\pi + \lambda_\nu, \, \mu_\pi + \mu_\nu)$ irrep will dominate
the low-lying energy structure.

In Table 2, we display the calculated  Gamow-Teller matrix elements, energy
denominators, predicted and experimentally determined double beta half lives of
$^{150}$Nd and $^{238}$U. They are given in the last column of Table 2 and
these half lives are lower bounds because the nuclear
matrix elements could be reduced by considering, for example the inclusion
of the pairing interaction, and therefore giving longer (never shorter)
$ \beta \beta$ half lives.

\vspace{2truepc}

\centerline{\large \bf REFERENCES}

\vspace{1truepc}

\small
\noindent
{1.} $ \ \ $K. T. Hecht and A. Adler, {\it Nucl. Phys.} {\bf A 137}, 129
(1969); \\ \hbox{~~~~~} R.D. Ratna Raju, J.P. Draayer and K.T. Hecht, {\it
Nucl. Phys.} {\bf A 202}, 433 \\  \hbox{~~~~~} (1973). \\
{2.} $ \ \ $A. Arima, M. Harvey and K. Shimizu, {\it Phys. Lett.} {\bf B 30},
517 (1969). \\
3.   $ \ \ $A. J. Kreiner, {\it Rev. Mex. F\ii s.} {\bf 40} Supl. 1, 22 (1994).
\\
4.   $ \ \ $F. S. Stephens et al., {\it Phys. Rev. Lett.} {\bf 64}, 2623
(1990). \\
5.  $ \ \ $R. Bergtsson, J. Dudek, W. Nazarewics and P. Olanders, {\it Physica
Scripta} {\bf 39}, \hbox{~~~~~} 196 (1989). \\
6.   $ \ \ $A. Bohr, I. Hamamoto and B. R. Mottelson, {\it Physica Scripta}
{\bf 26}, 267 (1982). \\
7.   $ \ \ $O. Casta\~nos , M. Moshinsky and C. Quesne, {\it Phys. Lett.} {\bf
B 277}, 238 (1992);\\ \hbox{~~~~~} O. Casta\~nos, V. Vel\'azquez, P.O. Hess and
J.G. Hirsch, {\it Phys. Lett.} {\bf B 321}, 303 \hbox{~~~~~} (1994). \\
8. $ \ \ $O. Casta\~nos , J. G. Hirsch, O. Civitarese and P. O. Hess, {\it
Nucl. Phys.} {\bf A571},  \hbox{~~~~~} 276 (1994). \\
9. $ \ \ \, $A. G\'o\'zd\'z, A. Staszczak and K. Zajac, {\it Acta Physica
Polonica.} {\bf B25},  665 (1994). \\
10. $ \, $D. Troltenier, W. Nazarewics, Z. Szymanski and J. P. Draayer, {\it
Nucl. Phys.} \hbox{~~~~~~}{\bf A567},  591 (1994). \\
11. $ \, $J. P. Draayer and K. J. Weeks, {\it Ann. Phys.} {\bf 156}, 41
(1984).\\
12.  $ \, $M. Doi, T.Kotani and E. Takasugi, {\it Progr. Theo. Phys. Suppl.}
{\bf 83}(1985) 1. \\
13.  $ \, $M. K. Moe, P. Vogel, {\it Ann. Rev. Nuc. Part. Sci.} (to be
published). \\
14.  $ \, $O. Civitarese and J. Suhonen,  {\it Phys. Rev.} {\bf C47} (1993)
2410. \\
15.  $ \, $O. Casta\~nos, J.P. Draayer and Y. Leschber, {\it Z.
Phys.} {\bf A 329} (1988) 33;\\ \hbox{~~~~~} H.A.Naqvi and J.P. Draayer,  {\it
Nucl. Phys.} {\bf A516} (1990) 351. \\
16.   $ \, $P. Ring, P. Schuck, {\it The Nuclear Many Body Problem}, (Springer
Verlag, New  \hbox{~~~~~} York 1980) .
\end{document}